\documentclass[journal=apchd5,manuscript=article,layout=twocolumn]{achemso}

\usepackage{achemso}
\usepackage{chemformula} 
\usepackage{jabbrv}
\usepackage{pdfpages}

\author{Miguel Y. Bacaoco}
\affiliation[UTS]
{School of Mathematical and Physical Sciences, University of Technology Sydney, Sydney, New South Wales, 2007, Australia}
\altaffiliation{Sydney Quantum Academy, Sydney, New South Wales, 2007, Australia}

\author{Kirill Koshelev}
\affiliation{Research School of Physics, Australian National University, Canberra, Australian Capital Territory, 2601, Australia}
\email{kirill.koshelev@anu.edu.au}

\author{Alexander S. Solntsev}
\affiliation[UTS]
{School of Mathematical and Physical Sciences, University of Technology Sydney, Sydney, New South Wales, 2007, Australia}
\altaffiliation{Sydney Quantum Academy, Sydney, New South Wales, 2007, Australia}
\email{alexander.solntsev@uts.edu.au}

\title[An \textsf{achemso} demo]
  {Third-Order Spontaneous Parametric Down Conversion in Dielectric Nonlinear Resonant Metasurfaces}

\abbreviations{IR,NMR,UV}

\begin{document}


\begin{abstract}

We propose a general scheme to investigate photon triplet generation (PTG) via third-order spontaneous parametric downconversion (TOSPDC) in $\chi^{(3)}$ nonlinear structures. Our approach leverages the quantum-classical correspondence between TOSPDC and its reverse classical process, three-wave sum-frequency generation (TSFG), to efficiently estimate the PTG rate. We apply this framework to nonlinear metasurfaces supporting quasi-bound states in the continuum (qBICs) in the optical range. From numerical analysis of non-collinear TSFG with degenerate input waves at qBIC wavelengths, we predict wavelength-tunable three-photon emission with spatio-angular correlations. These findings establish a novel method for modelling TOSPDC and also highlight the potential of nonlinear resonant metasurfaces as compact free-space photon triplet sources with quantum state control.

\end{abstract}

\section*{Introduction}

Quantum photonics is at the forefront of scaling quantum technologies, providing compact, scalable, and robust platforms for quantum computing, communication, and sensing~\cite{Wang2020,OBrien2009,Moody2022}. The key platform that enable this progress are nonlinear optical processes, which offer powerful tools for generating and manipulating non-classical states of light, such as single photons, squeezed states, and entangled photons~\cite{Caspani2017,Moody2020}.
\cite{Wang2020}
Among these processes, spontaneous parametric downconversion (SPDC) emerged as a cornerstone in quantum photonics, routinely used in $\chi^{(2)}$ nonlinear materials to produce correlated and highly entangled photon pairs~\cite{Wang2021}. These photon pairs have been instrumental in enabling quantum technologies, yet advanced applications demand more complex quantum states involving higher-order correlations. For instance, photon triplets, offer unique advantages such as tripartite entanglement and non-Gaussian quantum statistics~\cite{Agusti2020,Chekhova2005,Bencheikh2022}, which are critical for advancing quantum networks~\cite{Hillery1999}, error-resilient quantum computing~\cite{Braunstein2005,Takahashi2010,Weedbrook2012,Zhang2021}, and multiphoton quantum interference experiments~\cite{Banaszek1997,Agne2017}. 

The conventional method to produce photon triplet states is by third-order SPDC (TOSPDC) in $\chi^{(3)}$ nonlinear materials. However, such approach faces significant challenges primarily due to the inherently low third-order nonlinearity of most crystals and the stringent phase-matching requirements~\cite{Boyd2008,Bencheikh2007}. While both theoretical and experimental efforts have explored various platforms for TOSPDC, including bulk crystals \cite{Bencheikh2007,Borshchevskaya2015}, waveguides \cite{Moebius2016,Akbari2016,Huang2018,Banic2022a,Banic2022b}, and hybrid optical fibers\cite{Corona2011a,Corona2011b,Cavanna2016,Cavanna2020}, success has been limited due to material constraints and phase-matching challenges.  Given these difficulties, alternative approaches for generating photon triplets have been explored, such as cascaded second-order SPDC~\cite{Hubel2010,Shalm2013,Hamel2014,Krapick2016} and excitonic transitions in quantum dots~\cite{Khoshnegar2017}. However, cascaded SPDC scales as $|\chi^{(2)}|^2$ and inherently lacks the non-Gaussian properties required for advanced applications, while quantum-dot-based methods suffer from limited control over the downconverted frequencies and significant optical losses. To date, photon triplets via TOSPDC have only been demonstrated in the microwave regime using a superconducting parametric cavity~\cite{Chang2020}. Thus, the realization of optical TOSPDC remains an open problem, presenting both as a significant challenge and an exciting opportunity for advancing quantum photonics and multipartite entanglement.

Addressing this challenge requires innovative platforms that enhance nonlinear interactions and relax phase-matching constraints, and nonlinear dielectric resonant metasurfaces have emerged as promising candidates for this purpose~\cite{Okoth2019_nonphasematched,Solntsev2021,Sharapova2023,Vabishchevich2023}. Such metasurfaces support high-quality resonances that significantly improve light-matter interactions~\cite{Koshelev2018,Koshelev2019}, boosting the efficiency of classical and quantum nonlinear processes such as harmonic generation~\cite{Anthur2020,Bernhardt2020,Zalogina2023} and second-order SPDC~\cite{Marino2019,Santiago-Cruz2021,Santiago-Cruz2022,Zhang2022}. Furthermore, their subwavelength, two-dimensional geometry eases the longitudinal phase-matching requirements~\cite{Okoth2019_nonphasematched}, overcoming the limitations associated with bulk and waveguide-based approaches, while also offering a wide range of tunability in the emission~\cite{Parry2021,Mazzanti2022}.

In this letter, we propose a general scheme for investigating photon triplet generation (PTG) via TOSPDC in $\chi^{(3)}$ nonlinear structures. We extend the quantum-classical correspondence between SPDC and its reverse classical process, sum-frequency generation (SFG) to third-order processes, and leveraged such correspondence to study TOSPDC via classical three-wave SFG (TSFG). We apply this approach to a nonlinear metasurface supporting quasi-bound states in the continuum (qBIC) designed for TSFG. By numerical methods, we predict the PTG rates and emission profiles from the metasurface. This work demonstrates a novel approach for modelling TOSPDC, paving the way for new opportunities in tripartite quantum photonics.

\section*{Methods}
\subsection*{Third-Order Quantum Classical Correspondence}

Theoretical analysis of TOSPDC in compact geometric structures involves calculating the solution of the interaction Hamiltonian of form:~\cite{Banaszek1997,Chekhova2005}
\begin{equation}
    \hat{\rm H}(\mathbf{k,r}) = \hat{a}_1^\dagger \hat{a}_2^\dagger \hat{a}_3^\dagger \hat{b} \chi^{(3)}(\mathbf{r}) \Omega(\mathbf{k,r}) + \text{H. C.,}
\end{equation}
where $\hat{a}^\dagger$ is the bosonic creation operator for the output photons, that we label with indices 1, 2, and 3, $\hat{b}$ is the bosonic annihilation operator for the pump, $\chi^{(3)}(\mathbf{r})$ is the third-order susceptibility, $\Omega(\mathbf{k,r})$ accounts for the modal overlap, phase-matching, and other inhomogeneities, and H.C. stands for its Hermitian conjugate.  In the presence of anisotropy, losses, and dispersion, $\Omega$ becomes increasingly complex and the Hamiltonian also requires integration over both real and momentum spaces. This renders the first-principle calculation approach to be analytically and computationally demanding. 

For second-order SPDC, several semi-classical approaches have been developed to address the complexity of full quantum-mechanical treatment in various nonlinear media \cite{Schrder1983,Helt2012_SPDC,bertrand2025semiclassical,Kulkarni2022_SPDC-DFG,Poddubny2016}. For example, by performing classical measurements such as difference-frequency generation (DFG) or second-harmonic generation (SHG) in nonlinear waveguides with a low-intensity continuous-wave pump, it is possible to predict the average photon power in the SPDC regime \cite{Helt2012_SPDC}. However, at higher pump powers, the interaction becomes non-perturbative, and such models start to break down. Semi-classical models covering both low- and high- intensity pump power regimes have also been reported based on classical three-wave mixing which includes a pump and its coupling to a classical noise that simulates vacuum fluctuations \cite{Schrder1983,Kulkarni2022_SPDC-DFG,bertrand2025semiclassical}. Other approaches model SPDC wavefunctions and correlations as paraxial propagation of a
classical partially-coherent source \cite{DiLorenzoPires2011} and as linear propagation of classical light in coupled waveguide arrays \cite{Grafe2012}. Though most models were successful in predicting and explaining experimental measurements, they are often limited to specific structures, pump configurations, and phase-matching conditions, and typically do not account for absorption, scattering, and other physical real-world sources of inhomogeneities. A more general black-box approach addressed these limitations by developing an exact relationship between SFG efficiency and SPDC biphoton emission rates, utilizing Green's function formalism and Lorentz reciprocity principle valid for any arbitrary $\chi^{(2)}$ nonlinear structures \cite{Poddubny2016,Poddubny2020}. While such method have proven valuable in the second-order regime, its direct extension to third-order processes like TOSPDC has remained unexplored. 

As TOSPDC has not yet been realized in the optical domain, previous works have relied on its reverse classical degenerate process, third-harmonic generation (THG), for experimental validation and structural optimization~\cite{Richard2011,Moebius2016,Cavanna2016,Cavanna2020}. However, most of the calculation of PTG rates are still dependent on rigorous quantum-mechanical or semi-classical treatments~\cite{Dominguez-Serna2020,Banic2022a,Okoth2019} specific to the structure under study. 

This motivated our investigation, which aims to provide a general black-box alternative and an efficient method to study TOSPDC through its reverse classical process, TSFG, covering both degenerate (THG) and non-degenerate regimes. In this section, we first establish the quantum-classical correspondence between the third-order nonlinear processes: TOSPDC and TSFG. This approach is a higher-order extension of the previously established second-order correspondence\cite{Poddubny2016,Poddubny2020}, which has been successfully applied in studying photon-pair generation in $\chi^{(2)}$ nonlinear media~\cite{Marino2019,Santiago-Cruz2021,Parry2023,Zhang2022}. 

In general, TOSPDC describes the spontaneous decay of a pump photon with energy $\hbar \omega_p$ to three lower-energy photons with energies $\hbar \omega_1$, $\hbar \omega_2$, and $\hbar \omega_3$, obeying the energy conservation law $\omega_p = \omega_1 + \omega_2 + \omega_3$, and momentum conservation law (phase-matching) $\mathbf{k}_p = \mathbf{k}_1 + \mathbf{k}_2 + \mathbf{k}_3$. Conversely, TSFG involves the nonlinear sum-frequency mixing of three waves with frequencies $\omega_i$ and wavevectors $\mathbf{k}_i$ ($i=1,2,3$) generating a photon with frequency $\omega_p$, while obeying the same conservation laws as TOSPDC. A schematic representation of these processes is shown in Figure 1.

\begin{figure}[tb!]
    \centering
    \includegraphics[width=\linewidth]{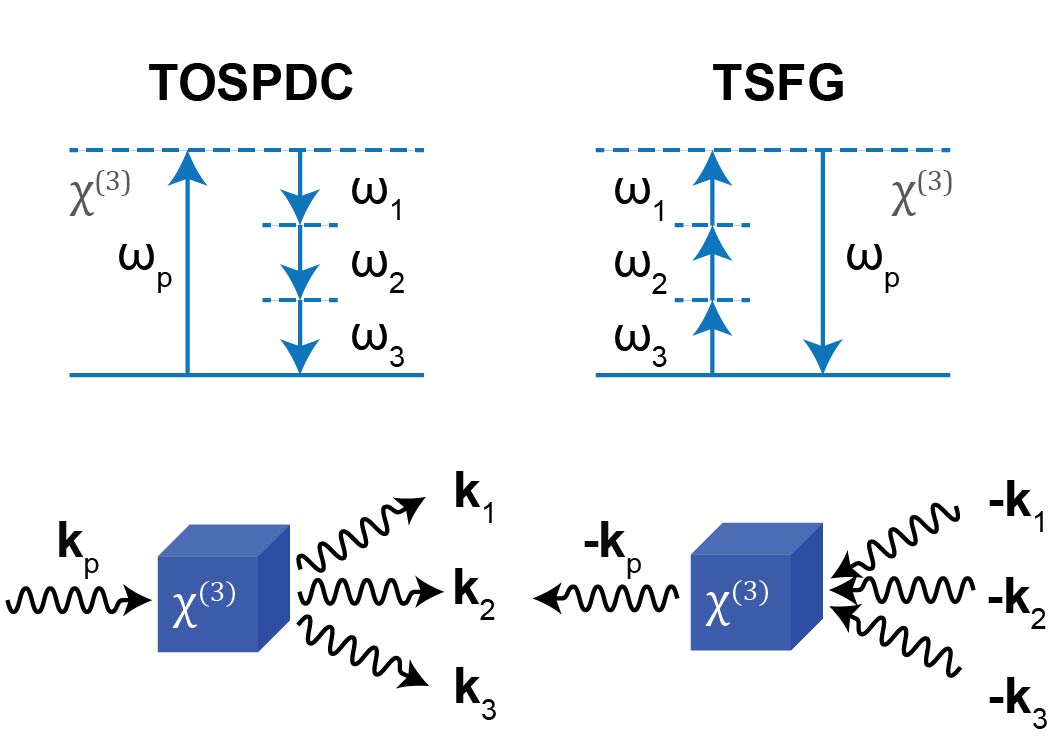}
    \caption{{\bf Quantum-classical correspondence concept.} Schematic representation of TOSPDC and TSFG in a $\chi^{(3)}$ nonlinear medium.}
    \label{fig:correspondence}
\end{figure}

Using Green's function formalism in Gaussian unit system~\cite{Poddubny2016,Poddubny2020}, we can write the photon triplet wavefunction $\Psi$, characterized by each photon's position $\mathbf{r_i}$, polarization $\sigma_i$, and frequency $\omega_i$, as:
\begin{multline}
	\Psi (\mathbf{r_1,r_2,r_3},\sigma_1, \sigma_2,\sigma_3,\omega_1,\omega_2,\omega_3) = \\ \int \text{d$^3$r$_0$} \sum_{\alpha, \beta, \gamma, \delta} G_{\sigma_1 \alpha}(\mathbf{r_1,r_0};\omega_1) G_{\sigma_2 \beta}(\mathbf{r_2,r_0};\omega_2) \\ \times G_{\sigma_3 \gamma}(\mathbf{r_3,r_0};\omega_3) \chi^{(3)}_{\alpha \beta \gamma \delta} (\mathbf{r_0}) E_{p \delta}(\mathbf{r_0}).
	\label{eqn:tospdc_wfcn}
\end{multline}
Similarly, the TSFG wavefunction $E_{p}(\mathbf{r_p},\sigma_p,\omega_p)$ can be written as:
\begin{multline}
	E_{TSFG,\sigma_p}(\mathbf{r_p},\omega_p) = \int\text{d$^3$r$_0$} \sum_{\alpha, \beta, \gamma, \delta} G_{\sigma_p \delta}(\mathbf{r_p,r_0};\omega_p) \\ \chi^{(3)}_{\alpha \beta \gamma \delta} (\mathbf{r_0}) E_{1 \alpha} (\mathbf{r_0}) E_{2 \beta} (\mathbf{r_0}) E_{3 \gamma} (\mathbf{r_0}).
	\label{eqn:tsfg_field}
\end{multline}
Here, $G$ is the electromagnetic Green's function. The subscripts $\sigma_i, \alpha, \beta,\gamma, \delta = x,y,z$ denote Cartesian indices. The vectors $\mathbf{r}_i$ are the far-field positions parallel to $\mathbf{k}_i$ where photon mode $i$ is detected for the TOSPDC process, while in the TSFG process, $\mathbf{r}_i$ refer to the location of point dipoles generating the plane wave $-\bf{k}_i$. By writing the classical fields $E_{i \alpha}$ in terms of the polarization $P_{i\alpha}$ inducing such fields as: 
\begin{equation}
	E_{i \sigma_i}(r_i) = \int 	\text{d$^3$r$_0$} \sum_{\alpha} G_{\sigma_i \alpha}(\mathbf{r_i,r_0};\omega_i) P_{i\alpha}(r_0),
\end{equation}
and comparing Equations \ref{eqn:tospdc_wfcn} and \ref{eqn:tsfg_field}, we arrive at the general Lorentz reciprocity relation for third-order processes as: 
\begin{multline}
	\int \text{d$^3$ r$_1$} \int \text{d$^3$r$_2$} \int \text{d$^3$r$_3$} \Psi  \mathbf{P}_1\mathbf{(r_1)} \mathbf{P}_2\mathbf{(r_2)} \mathbf{P}_3\mathbf{(r_3)}  \\ = \int \text{d$^3$r$_p$} E_{TSFG} \mathbf{P}_p\mathbf{(r_p),}
	\label{eqn:correspondence}
\end{multline}
where the arguments of $\Psi$ and $E_{TSFG}$ were dropped for brevity. This relationship allows the expression of the three-photon wavefunction in terms of classical parameters $E_{TSFG}$ and $\mathbf{P}_i \mathbf{(r_i)}$. We then calculate the transition rate of the parametric interaction using Fermi's golden rule as: 
\begin{multline}
W_{123} = \frac{2 \pi}{\hbar} \delta(\hbar \omega_p - \hbar \omega_1 - \hbar \omega_2 - \hbar \omega_3) \\ \times \left| \sum_{m,n,o} d^*_{1m} d^*_{2n} d^*_{3o}  \Psi \right|^2,
	\label{eqn:3phtranrate}
\end{multline}	
in idealized  photodetectors with dipole moments $\mathbf{d_i^*}$ whose Cartesian components are indexed by $m,n,o=x,y,z$. We get the experimentally observable PTG rate by integrating over the spectral width of idler photons $2$ and $3$ and across the half-space where the photons are propagating (See Supporting Information).

This leads to the central result of this section, which establishes a direct relationship between the PTG rate and the efficiency of the classical TSFG process as:
\begin{multline}
    \frac{dN_{\text{triplet}}}{dt d\Omega_1 d\Omega_2 d\Omega_3} = \frac{\hbar c}{2 \pi} \frac{\lambda_p^4}{\lambda_1^3 \lambda_2^3 \lambda_3^3}  \Phi_p \\ \times \iiint d\omega_1 d\omega_2 d\omega_3 \delta(\omega_p-\omega_1-\omega_2-\omega_3) \frac{d \Xi^{\text{TSFG}}}{d\Omega_p}
    \label{eqn:ptg}
\end{multline}

where $\frac{d \Xi^{\text{TSFG}}}{d\Omega_p}$ is the classical differential TSFG efficiency defined by the ratio of the output TSFG flux $\Phi_{TSFG}$ to the input fluxes $\Phi_i$ of the interacting waves over the cross section $A$:
\begin{equation}
 	\frac{d \Xi^{\text{TSFG}}}{d\Omega_p}= \frac{\Phi_{TSFG}}{\Phi_1 \Phi_2 \Phi_3}A.
\end{equation}

We note the generality of Equation \ref{eqn:ptg} which is exact in the absence of other nonlinear processes and valid for any arbitrary $\chi^{(3)}$ structures. It establishes a "black-box" approach that quantitatively predicts the TOSPDC rate while inherently accounting for effective field enhancement, dispersion, and losses in the structure. Most importantly, it provides an efficient method to compute and optimize PTG through a more accessible classical TSFG experiment. This advancement simplifies the experimental validation and also enables practical implementations of TOSPDC in complex photonic architectures.

\section*{Results \& Discussion}
    
\subsection*{Nonlinear Resonant Metasurface Design}

\begin{figure*}[tb!]
    \centering
    \includegraphics[width=\linewidth]{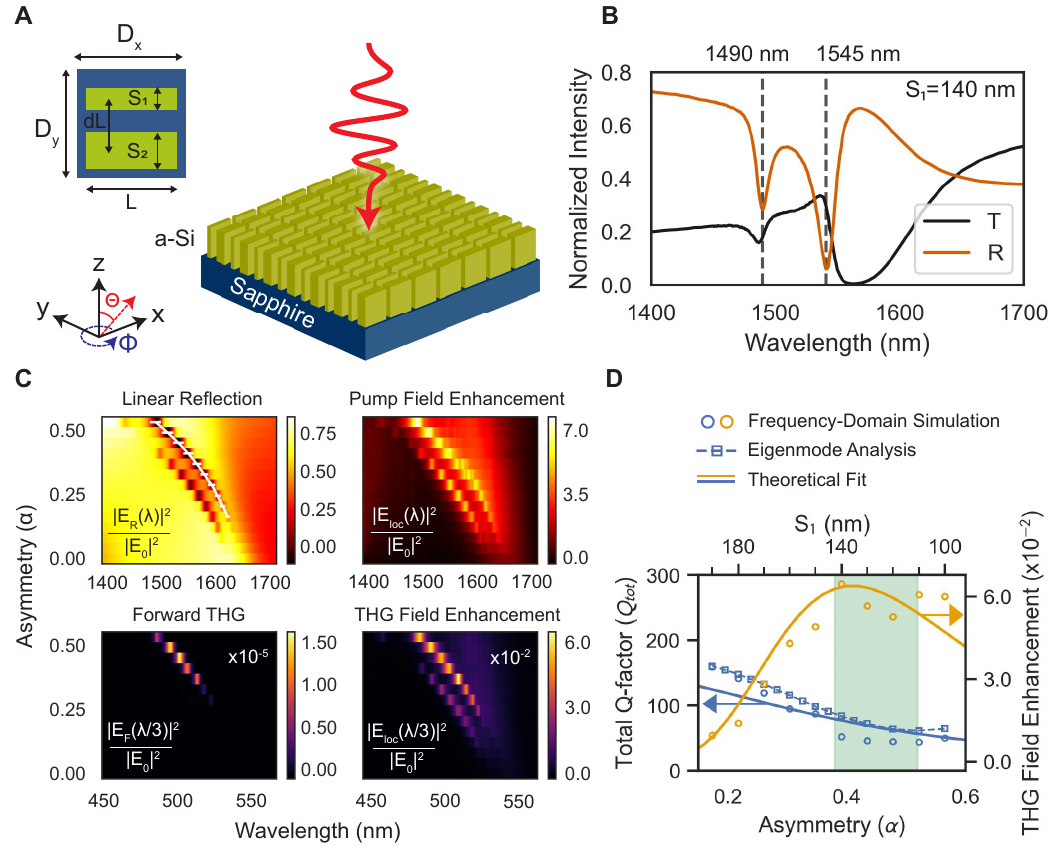}
    \caption{\textbf{Classical optical properties of the metasurface.} \textbf{(A)} Schematic representation of the metasurface with a square unit cell consisting of two parallel bars with different widths, excited by a normally incident plane wave. \textbf{(B)} The calculated spectra in the transmission, T, and reflection, R, when $S_1=140$~nm and $S_2=230$~nm corresponding to asymmetry $\alpha=0.44$. The qBIC is observed at 1545nm. \textbf{(C)} Linear and nonlinear spectral response at various asymmetry values ranging from 0 to 0.55. The far-field response is displayed in the first column while the associated near-field enhancement is on the second column. \textbf{(D)} The calculated total Q-factor and TSFG enhancement at varying asymmetry. The green-shaded region indicates the critical coupling regime. }
    \label{fig:meta_classical}
\end{figure*}

We now apply the general theory to a nonlinear dielectric metasurface structure supporting optical qBICs. We adopt a qBIC structure consisting of two rectangular silicon bars with equal length and height but with varying widths on a sapphire substrate and arranged in a square lattice. This geometry is known to support high-quality qBIC modes and has been extensively studied in various nonlinear optical processes \cite{Koshelev2019,Bernhardt2020,Sinev2021,Zograf2022}. The structure is engineered to support qBIC mode in the telecommunications spectral range (1400~nm - 1700~nm) by setting the bar length to $L=570$~nm and height to $H=615$~nm, with a center-to-center separation distance of $dL=330$~nm (Figure 2A) \cite{Koshelev2022}. This spectral range corresponds to the wavelengths of the output photons from degenerate TOSPDC pumped in the visible range. To induce the qBIC, the bars must have different widths $S_1$ and $S_2$ \cite{Koshelev2018}, one is fixed at $S_2=230$~nm and the other is varied such that $S_1<S_2$. The meta-atom is arranged in a periodic lattice along both the $x$ and $y$ directions with a periodicity of $D_x=D_y=680$~nm. 

Guided by the quantum-classical correspondence, the structure is first optimized for TSFG process. Symmetry-protected BICs can be engineered to achieve critical coupling, where the nonlinear enhancement remains maximal even in the presence of losses, by adjusting the asymmetry, $\alpha (S_1) =(S_2 - S_1)/S_2$, of the bar widths~\cite{Koshelev2018}. Due to the interplay of the resonance quality factor, nonlinear field enhancement, asymmetry, and losses \cite{Koshelev2018,Koshelev2019}, this step is crucial to achieve high-performance resonators for maximum TSFG and correspondingly, TOSDPC. Non-radiative sources of loss such as material imperfections, fabrication defects, and surface roughness are emulated to have an effective non-radiative quality factor equal to $Q_{nr}=150$, which has been experimentally verified for Si qBIC metasurfaces in the near-infrared range~\cite{Kuhne2021}. Dispersion and absorption are also taken into account from the material's complex refractive index obtained from ellipsometric measurement and from literature~\cite{Malitson1962,Aspnes1983}. 

Full-field frequency-domain electromagnetic simulations were performed using a normally incident TM (x-polarized) plane wave as the pump, with varying asymmetry and wavelength, through finite-element modeling (See Supporting Information). Both linear and nonlinear responses were calculated separately. In general, the nonlinear wave-mixing process is governed by the $\chi^{(3)}$ tensor and the Cartesian components of the interacting waves $\mathbf{E}(\omega_i)$, expressed as \cite{Boyd2008}
\begin{equation}
    P_i^{(3)} (\omega_p) = 6\epsilon_0 \sum_{j,k,l}  \chi_{ijkl}^{(3)} E^{}_{j}(\omega_1) E_{k} (\omega_2) E_{l} (\omega_3), 
\end{equation}
where $P^{(3)}$ is the third-order nonlinear polarizability, and $i,j,k,l = x,y,z$ denote the Cartesian coordinate indices. This formulation is implemented numerically, and the transverse components of the zeroth diffraction-order TSFG field are calculated. In designing the metasurface, three degenerate collinear input pumps with equal intensities $|E_0|^2$ were considered to generate the nonlinear field. In such a case, the TSFG process becomes equivalent to THG. 

Several metrics are then determined to characterize the metasurface's optical response. The normalized intensities for the linear far-field transmission $T=|E_T|^2/|E_0|^2$ and reflection $R=|E_R|^2/|E_0|^2$ are calculated as the average $|E|^2/|E_0|^2$ along the bottom and top planes of the simulation cell, respectively. Similar analysis was also performed to calculate the respective forward and backward THG intensities. To analyze the near-field response in both linear and nonlinear regimes, the local field enhancement $|E_{\text{loc}}|^2/|E_0|^2$ was calculated as the maximum $|E|^2/|E_0|^2$ within the unit cell corresponding to a single point in the mode volume \cite{Maier2007}.

Figure 2B shows the linear transmission and reflection spectra at $S_1=140$~nm with a qBIC resonance at 1545~nm, characterized by their Fano-type lineshapes \cite{Koshelev2019}. The spectral map of the reflection for asymmetry parameter in the range from 0 to 0.55 is shown in the upper-left panel of Figure 2C. The qBIC mode dispersion obtained via eigenmode numerical calculations is shown with white solid lines overlaid with the map, where the error bars indicate the mode bandwidth. We observe solid agreement of eigenmode and scattering calculations. 

The local field enhancement at the pump wavelength is also shown at the upper-right panel of Figure 2C. In the figure, we observe that the maximal enhancement occurs at the qBIC modes. This distribution is expected, as the asymmetric bars induce a net magnetic dipole in the meta-atom in the linear regime \cite{Koshelev2019,Bernhardt2020,Sinev2021,Zograf2022}, which effectively improves the confinement of the pump field.

We next analyze the nonlinear THG signal in the near- and far-field, using the frequency domain simulations in the undepleted pump approximation. The generated THG field in the forward direction $|E_{F}(\lambda/3)|^2$ and the associated near-field enhancement at third-harmonic wavelength $|E_{\text{loc}}(\lambda/3)|^2$ are shown in the lower-left and lower-right panels of Figure 2C, respectively. We can see that both spectra are clearly dominated by the qBIC mode, and the THG enhancement closely resembles the linear local field enhancement map.  Such similar map is the result of improved light confinement in the pump which consequently boosts the nonlinear response. Here, only the forward THG field is shown, as its intensity is an order of magnitude higher than the backward THG field, although both fields exhibit similar lineshapes (See Supporting Information). 

However, there is a non-trivial relationship between the maximum nonlinear response to asymmetry. As shown in Figure 2D, the local THG field enhancement at the resonant wavelength (orange circles) increases with asymmetry, reaching a peak before decreasing at higher asymmetry values ($\alpha \approx 0.8$ to 1, not shown), where it drops to $\approx 0.15$. This behavior can be explained by satisfying the critical coupling regime through asymmetry engineering. In this regime, the radiative ($Q_r$) and non-radiative ($Q_{nr}$) quality factors are almost equal, leading to optimal energy exchange between radiative and non-radiative channels in the resonator \cite{Koshelev2019}. The orange solid line shows the theoretical fit predicting a critical asymmetry (where $Q_r=Q_{nr}$) at $\alpha_{cr}=0.41$, which shows excellent agreement with the simulation.

Additionally, Figure 2D shows the evolution of the total qBIC quality factor $Q_{tot}=(Q_r^{-1}+Q_{nr}^{-1})^{-1}$, obtained from the frequency domain simulation (blue circles) and eigenmode analysis (blue squares) for different values of asymmetry. Theoretical fit (blue solid line) to the simulation results is also shown which qualitatively aligns well with the results. At zero asymmetry, the resonator behaves as a pure BIC with an infinite $Q_r$, and the stored energy in the system dissipates solely through non-radiative channels governed by $Q_{nr}$. From the fitting, the predicted total Q-factor at $\alpha=0$ is 146, closely matching the artificially introduced value of 150. Overall, from Figure 2D, we can see that the critical coupling regime is achieved at $\alpha \approx 0.4-0.5$, which corresponds to optimal $S_1 \approx 140-110$~nm (green shaded region). In this regime, the quality factor of the metasurface is approximately 50, and the far-field forward THG intensity reaches four orders of magnitude compared to an unstructured Si film of the same thickness (See Supporting Information).

\subsection*{TOSPDC Emission \& Rate Characterization}

Next, we study the three-photon emission using the optimized metasurface ($S_1=140$~nm) geometry and by utilizing the quantum-classical correspondence. Unlike the second-order SPDC where the emitted photon pairs are emitted in two conjugate angles according to momentum conservation (phase matching condition), the three-photon emission process in TOSPDC allows infinitely many wavevector configurations that satisfy this condition~\cite{Chekhova2005,Okoth2019}. Consequently, the photon triplet generated by TOSDPC are emitted over a broad continuum of free-space modes with different wavevectors. This motivated our investigation to consider the most symmetric emission configuration where the generated photon triplets are emitted with $C_3$ symmetric divergence shown in Figure 3A. In this case, each photon propagates at an angle $\theta$ from the pump propagation axis ($+z$), with their azimuthal directions separated by 120$^\circ$. We rotate this configuration along the $z$-axis via the azimuthal angle $\phi$ and through $\theta$ to characterize the angular and radial emission profiles, respectively. With this configuration, we note that the emission is fully collinear when $\theta=0$.

\begin{figure*}[ht!]
    \centering
    \includegraphics[width=\linewidth]{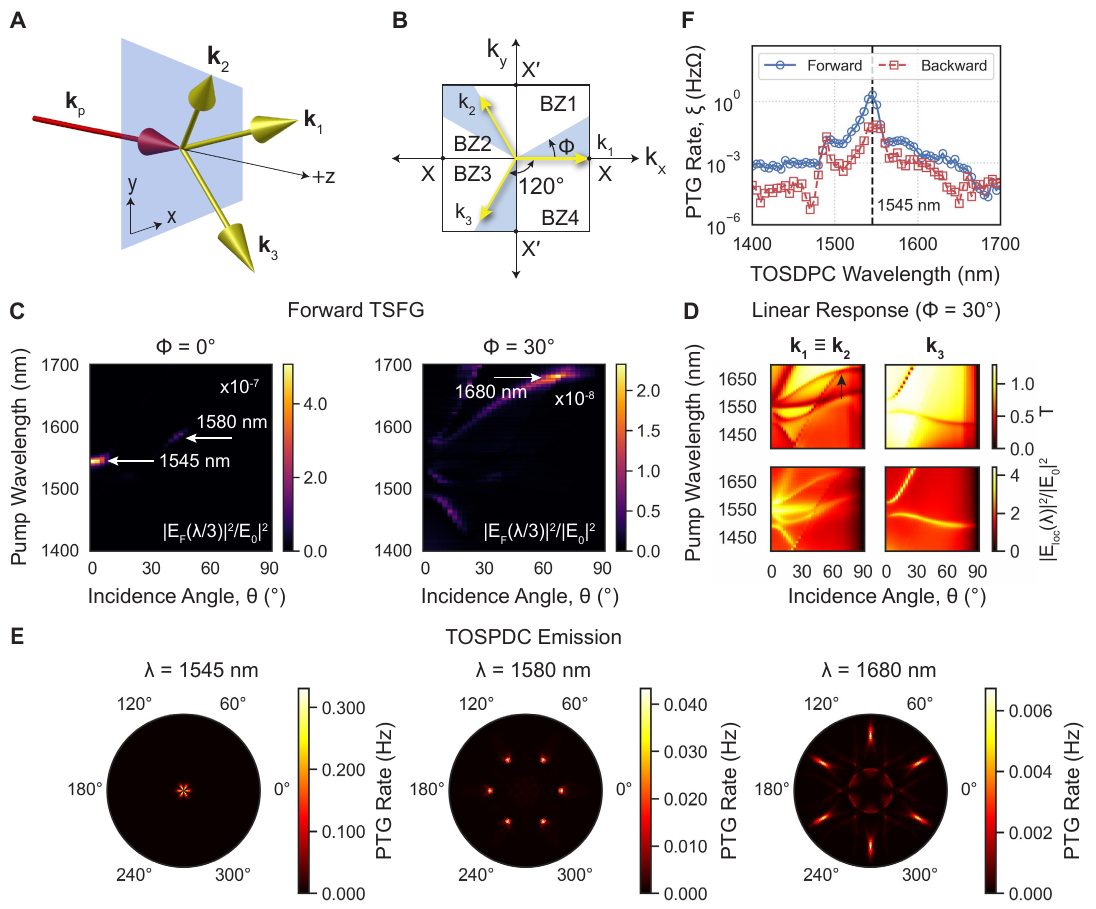}
    \caption{\textbf{TOSPDC emission modelling via classical non-collinear TSFG.} \textbf{(A)} The general wavevector configuration for TOSPDC where the pump $\mathbf{k_p}$ produces photon triplets with wavevectors $\mathbf{k_i},$ $ (i=1,2,3)$. \textbf{(B)} The k-space lattice of the metasurface with $C_2$ overall symmetry containing four Brillouin zones (BZs). The three-fold symmetric photon emission is overlaid as yellow arrows, and rotating such configuration from 0-30$^\circ$ allows the mapping of all unique regions of the BZ (blue-shaded regions) due to the meta-atom symmetry and the photons being degenerate. \textbf{(C)} The nonlinear far-field spectral response from degenerate non-collinear TSFG simulation at $\phi=0^\circ$ and $\phi=30^\circ$. A few resonant points are identified at $\lambda=1545$~nm, $\lambda=1580$~nm, $\lambda=1680$~nm. \textbf{D} The linear optical response of non-collinear input waves $\mathbf{k_i}$ in the far-field (T) and near-field $E_{\text{loc}}$ at $\phi=30^{\circ}$. \textbf{(E)} The constructed TOSPDC emission profile calculated from the TSFG maps via the quantum classical correspondence at the identified resonance points. \textbf{(F)} The integrated PTG rate across all emission angles at different pump wavelengths.}
    \label{fig:meta_quantum_v3.3}
\end{figure*}

Moreover, the $C_2$ symmetric meta-atom embedded in a square lattice with $C_4$ symmetry gives the system an overall $C_2$ symmetry with a square Brillouin zone (BZ). Figure 3B shows the k-space lattice with four BZs. By considering the assumed wavevector configuration, illustrated as yellow arrows in the figure, we can cover all the unique areas of the BZ by only rotating the configuration from 0 to 30 degrees. These are schematically represented by the blue-shaded regions in Figure 3B. 

We model this TOSPDC configuration by considering its reverse classical counterpart, noncollinear TSFG, where three propagating linearly polarized waves $\mathbf{E}(\omega_i, \mathbf{k}_i)_{i=(1,2,3)}$ interact within the material and generate the TSFG field (See Supporting Information). 

Figure 3C shows the resulting TSFG field for $\phi=0^\circ$ and $\phi=30^\circ$ across the polar range $\theta=0-90^\circ$. Under near collinear excitation, the qBIC resonance at the pump (1545nm) dominates the nonlinear enhancement in the TSFG regime, which confirms the result of the structure optimization performed at normal incidence. However, at azimuth angle $\phi=30^\circ$, another high-quality resonance emerges at a different wavelength, excited at an oblique polar angle of approximately $\theta \approx 60^\circ$. This enhancement originates from the modal interference among the three independent input waves incident at different oblique angles. Figure 3D shows the transmission (upper panels) and local field enhancement (lower panels) associated with these three input waves at $\phi=30^\circ$. Due to symmetry (see Figure 3B), the responses from $\mathbf{k_1}$ and $\mathbf{k_2}$ are equivalent. The transmission profile reveals that the TSFG resonance at $\theta \approx60^\circ$ appears to originate from the crossing of two resonant features excited at higher polar angles. Furthermore, the additive effect of the two symmetric excitations further enhances the light confinement, contributing to the observed nonlinear response. At the same polar angle, $\mathbf{k_3}$ doesn't seem to excite any modes. Looking at the reverse process, this result suggests the feasibility of resonantly emitting TOSPDC photon triplets in different directions with oblique wavevectors $\mathbf{k}_1$, $\mathbf{k}_2$, and $\mathbf{k}_3$ as in Figure 3A.

We then predict the TOSPDC emission profile from the TSFG maps using the quantum classical correspondence introduced in Equation 7. Figure 3E shows the constructed spatial distribution of the calculated emission rate at TOSPDC wavelengths $\lambda=1545$~nm, $\lambda=1580$~nm, and $\lambda=1680$~nm. Here, we assumed the pump power to be $\Phi_p=40$~GWcm$^{-2}$ focused to a spot size with radius $r=1$~um and the bandwidth $\Delta \lambda=10$~nm (See Supporting Information). 

At the qBIC resonance ($\lambda=1545$~nm), the emission remains predominantly collinear, with divergence of less than 10$^\circ$, as shown in the leftmost panel of Figure 3E. This enables efficient collection of all emitted photon triplets using low numerical aperture (NA) objective for various applications. 

Interestingly, by switching the pump wavelength to excite the oblique modes ($\lambda=1580$~nm and $\lambda=1680$~nm), TOSPDC photons can be emitted non-collinearly, as shown in the center and rightmost panels of Figure 3E. In this configuration, the angular freedom inherent in the TOSPDC process allows the simultaneous excitation of all oblique modes, resulting in photons emitted into six distinct lobes with substantial radial separation. At $\lambda=1580$~nm, the individual emission is narrow and has a radial distance of approximately 45$^\circ$. A similar emission profile can be obtained at $\lambda=1680$~nm, where the lobes have a larger radial distance of about 60$^\circ$ with highly elongated shape. These lobes define a spatio-angular basis for the photon triplets, which can be separately collected by different low-NA objectives in the far-field, while preserving their quantum correlations.

However, the $C_3$ symmetric momentum conservation configuration considered in this study (Figure 3A-B) should be considered in interpreting these spatio-angular bases. This means that if the first photon is measured at an azimuthal angle of $\phi_1=30n^{\circ}$ where $n \in \{0,1,2,...,11\}$, the model predicts that the other two photons will be found at $\phi_2=(30n+120)^{\circ}$ and $\phi_3=(30n-120)^{\circ}$. While other asymmetric wave-vector configurations satisfying momentum conservation are possible, their analysis merits further investigation in future research. Overall, these results introduce a novel approach for generating three-photon correlations, which could be valuable for future multiphoton applications.

By integrating over all emission angles in the half-sphere, we estimate the total PTG rate $\xi$ as:
\begin{equation}
    \xi = \Phi_p \iint \frac{dN_{\text{triplet}}(\theta,\phi)}{dt}  \sin{(\phi)} d\phi d\theta.
\end{equation}

Figure 3F shows the calculated forward and backward photon triplet emission rates at all other pump wavelengths and  fixed pump power $\Phi_p=40~$GWcm$^{-2}$. At $\lambda_p=515$~nm, photon triplets can be collected at a rate of 2 Hz at 3$\lambda_p=1545$~nm, a rate well within the capability of state-of-the-art low-noise single-photon detectors. This value further scales by broadening the resonance bandwidth via metasurface engineering and increasing the pump power. Overall, these results provide a benchmark for such a simple configuration, potentially enabling the eventual physical realization of TOSPDC in the optical domain.  

\section*{Conclusion \& Outlook}

In conclusion, we have developed a practical approach for studying photon triplet states and demonstrated its application to nonlinear resonant metasurface platform. By establishing the quantum-classical correspondence between TOSPDC and TSFG, we provided a quantitative estimate of the photon triplet emission based on the more experimentally accessible TSFG efficiency. We then designed a qBIC metasurface optimized for TSFG at normal incidence, incorporating noise emulation to reflect real-world conditions. Using the reverse non-collinear TSFG simulations, we explored the TOSDPC emission characteristics in the most symmetric wavevector configuration, where all three photons are equidistant in the azimuthal and radial directions. Furthermore, our results revealed a tunable emission profile, arising from modes excited at various k-locations, which can be switched by adjusting the pump wavelength.

Looking ahead, we anticipate the experimental verification of the presented theories and numerical findings. In addition, we foresee the adoption of our approach using the third-order correspondence and nonlinear resonant metasurfaces in quantum photonics, particularly in the generation and manipulation of multi-photon states for advanced quantum technologies. However, several practical challenges remain underexplored. Although the combined effects of fabrication defects and material losses are modeled using a phenomenological loss parameter ($Q_{nr}$) in the simulations, it remains crucial to experimentally verify this value and to investigate the individual contributions of each factors. Thermal effects in metasurfaces are also an emerging area of study \cite{Lee2025-ny}, with the potential to offer additional control over resonator behavior. Addressing these factors will be essential for guiding future experimental implementations and for realizing robust, scalable quantum photonic platforms.

\section*{Associated Content}

\subsection*{Supporting Information}

The following file is available free of charge.
\begin{itemize}
  \item Detailed formulation of the TOSPDC-TSFG correspondence; Details of the numerical simulation; Additional classical simulation results; Details of TOSPDC rate calculation
\end{itemize}

\section*{Author Information}

\subsection*{Corresponding Author}
Alexander S. Solntsev -- School of Mathematical and Physical Sciences, University of Technology Sydney, Sydney, New South Wales, 2007, Australia; alexander.solntsev@uts.edu.au

\noindent
Kirill Koshelev --
Research School of Physics, Australian National University, Canberra, Australian Capital Territory, 2601, Australia; kirill.koshelev@anu.edu.au

\subsection*{Funding Sources}

 K.K. acknowledges support from the Australian Research Council (ARC) through the Discovery Early Career Researcher Award DE250100419.

\subsection*{Notes}

The authors declare no conflicts of interest.

\section*{Acknowledgment}

M.B. and A.S. acknowledge the support from Sydney Quantum Academy. M.B. acknowledges UTS Faculty of Science for financial support with the International Research Scholarship.


\bibliography{achemso_demo}
\newpage


\section*{Supplementary material}

# Supporting Information: Third-Order Spontaneous Parametric Down Conversion in Dielectric Nonlinear Resonant Metasurfaces

Miguel Y. Bacaoco,[†,¶] Kirill Koshelev,[*,‡] and Alexander S. Solntsev[*,†,¶]

†School of Mathematical and Physical Sciences, University of Technology Sydney, NSW 2007, Australia

‡Research School of Physics, Australian National University, Canberra ACT 2601, Australia

¶Sydney Quantum Academy, NSW, Australia

E-mail: kirill.koshelev@anu.edu.au; alexander.solntsev@uts.edu.au

## 1  Derivation of the third-order quantum-classical correspondence

This section derives the quantum-classical correspondence between third-order spontaneous parametric downconversion (TOSPDC) and three-wave sum-frequency generation (TSFG), that was used to calculate the photon-triplet generation (PTG) rate from classically accessible TSFG efficiency. This builds from earlier works on 2nd order SPDC-SFG correspondence, which is based on the generalized Lorentz reciprocity and Green's function formalism.[S1,S2]

## 1.1 Generalized Lorentz Reciprocity Relation

In such a formalism,[S1] we can write the wavefunction of the photon triplet state with frequencies $\omega_i$ where $i = 1, 2, 3$ produced via TOSPDC from pump $\omega_p$ as:

$$\Psi(\mathbf{r_1}, \mathbf{r_2}, \mathbf{r_3}, \omega_1, \omega_2, \omega_3) = \int d^3 \, r_0 G_{\sigma_i\alpha}(\mathbf{r_1}, \mathbf{r_0}; \omega_1) G_{\sigma_2\beta}(\mathbf{r_2}, \mathbf{r_0}; \omega_2) G_{\sigma_3\gamma}(\mathbf{r_3}, \mathbf{r_0}; \omega_3)$$
$$\chi^{(3)}_{\alpha\beta\gamma\delta}(\mathbf{r_0}) E_{p\delta}(\mathbf{r_0}). \quad (1)$$

Whereas the nonlinear field $E_{TSFG}$ produced via three-wave sum-frequency generation at frequency $\omega_p = \omega_1 + \omega_2 + \omega_3$ in the nonlinear structure can be written as:

$$E_{TSFG}(\mathbf{r_p}, \omega_p) = \int d^3 \, r_0 G_{\sigma_p\delta}(\mathbf{r_p}, \mathbf{r_0}; \omega_p) \chi^{(3)}_{\alpha\beta\gamma\delta}(\mathbf{r_0}) E_{1\alpha}(\mathbf{r_0}) E_{2\beta}(\mathbf{r_0}) E_{3\gamma}(\mathbf{r_0}). \quad (2)$$

Here, the subscripts $\alpha, \beta, \gamma, \delta$ denote Cartesian indices, and $G_{\alpha\beta}$ is the electromagnetic Green's function sartisfying the differential equation:

$$\nabla \times \nabla \times G(\mathbf{r}, \mathbf{r_i}; \omega_i) = \left(\frac{\omega_i}{c}\right)^2 \epsilon(\mathbf{r}) G(\mathbf{r}, \mathbf{r_i}; \omega_i) + 4\pi \left(\frac{\omega_i}{c}\right)^2 \delta(\mathbf{r} - \mathbf{r_i}). \quad (3)$$

From Lorentz reciprocity principle, we introduce the polarizability densities $\mathbf{P_i}, (i = 1, 2, 3, p)$ inducing the corresponding fields $\mathbf{E_i}, (i = 1, 2, 3, p)$:

$$E_{i\alpha}(r) = \int d^3 \, r_0 G_{\sigma_i\alpha}(\mathbf{r_i}, \mathbf{r_0}; \omega_i) P_{i\alpha}(r_0), \quad i = 1, 2, 3, p. \quad (4)$$

This will then allow us to rewrite Equations 1 and 2 as:

$$\Psi(\mathbf{r_1}, \mathbf{r_2}, \mathbf{r_3}, \omega_1, \omega_2, \omega_3) = \int d^3 \, r_p \int d^3 \, r_0 G_{\sigma_i\alpha}(\mathbf{r_1}, \mathbf{r_0}; \omega_1) G_{\sigma_2\beta}(\mathbf{r_2}, \mathbf{r_0}; \omega_2) G_{\sigma_3\gamma}(\mathbf{r_3}, \mathbf{r_0}; \omega_3)$$
$$\chi^{(3)}_{\alpha\beta\gamma\delta}(\mathbf{r_0}) G_{\delta\sigma_p}(\mathbf{r_0}, \mathbf{r_p}; \omega_p) \mathbf{P_p}(\mathbf{r_p}) \quad (5)$$

and

$$E_{TSFG}(\mathbf{r_p};\omega_p) = \int d^3\,r_1 \int d^3\,r_2 \int d^3\,r_3 \int d^3\,r_0 G_{\sigma_p\delta}(\mathbf{r_p},\mathbf{r_0};\omega_p)$$

$$\chi^{(3)}_{\alpha\beta\gamma\delta}(\mathbf{r_0})G_{\alpha\sigma_i}(\mathbf{r_0},\mathbf{r_1};\omega_1)G_{\beta\sigma_2}(\mathbf{r_0},\mathbf{r_2};\omega_2)G_{\gamma\sigma_3}(\mathbf{r_0},\mathbf{r_3};\omega_3)\mathbf{P_1}(\mathbf{r_1})\mathbf{P_2}(\mathbf{r_2})\mathbf{P_3}(\mathbf{r_3}). \quad (6)$$

Comparing Equations 5 and 6 to the Lorentz reciprocity relation:

$$\int_V d^3r \mathbf{P_1}(\mathbf{r}) \cdot \mathbf{E_2}(\mathbf{r}) = \int_V d^3r \mathbf{P_2}(\mathbf{r}) \cdot \mathbf{E_1}(\mathbf{r}), \quad (7)$$

we arrive at the general reciprocity betweeen TOSPDC and TSFG:

$$\int d^3\,r_1 \int d^3\,r_2 \int d^3\,r_2 \Psi(\mathbf{r_1},\mathbf{r_2},\mathbf{r_3},\omega_1,\omega_2,\omega_3)\mathbf{P_1}(\mathbf{r_1})\mathbf{P_2}(\mathbf{r_2})\mathbf{P_3}(\mathbf{r_3})$$

$$= \int d^3\,r_p E_{TSFG}(\mathbf{r_p},\omega_p)\mathbf{P_p}(\mathbf{r_p}) \quad (8)$$

We will now use the established third-order quantum-classical correspondence to predict the photon-triplet generation rate in terms of classically accessible TSFG efficiency.

## 1.2 Differential TSFG Efficiency

We start by defining the electric field of the structure illuminated by three plane waves $\mathbf{E_i}$ ($i = 1, 2, 3$) with wavevectors $-\mathbf{k_i}$, schematically shown in Figure 1 of the main text. We then assume that the waves are generated by point dipoles $\mathbf{d_i^*}$ in the far-field located at $\mathbf{r_i} \parallel \mathbf{k_i}$, with unit amplitudes $\mathbf{d_i^*} \perp \mathbf{k_i}$. In such a case, the fields can be written as:

$$\mathbf{E_i}(\mathbf{r}) = G(\mathbf{r}, \mathbf{r_i}; \omega_i)\mathbf{d_i^*}. \quad (9)$$

From away from the structure where $\epsilon = 1$, the Green's function takes the form:

$$G_{\alpha,\beta}(\mathbf{r}, \mathbf{r_i}; \omega_i) = \left[\left(\frac{\omega_i}{c}\right)^2 + \frac{\partial}{\partial x_\alpha}\frac{\partial}{\partial x_\beta}\right]\frac{e^{i\omega_i|\mathbf{r}-\mathbf{r_i}|/c}}{|\mathbf{r} - \mathbf{r_i}|}. \tag{10}$$

By further assuming $r_i \gg r$, we can write the locally-plane waves $\mathbf{E_i}$ by substituting the Greens function definition above to Equation 9, which then takes the form:

$$\mathbf{E_i}(\mathbf{r}) = q_i^2 \frac{e^{iqr_i}}{r_i}\mathbf{d_i^*}, \tag{11}$$

where $q_i = \omega_i/c$. This then allows us to write the time-averaged photon flux (in Wm$^{-2}$), an experimentally accessible quantity, as:

$$\Phi_i = \frac{c}{2\pi}|E_i|^2 = \frac{c}{2\pi}\frac{q_i^4}{r_i^2}|d_i^*|^2. \tag{12}$$

The generated nonlinear TSFG field can then be calculated as a convolution of the illuminating fields and the Green's function at the TSFG frequency as:

$$E_\alpha^{TSFG}(\mathbf{r}_p \leftarrow \mathbf{r}_1, \mathbf{d}_1^*; \mathbf{r}_2, \mathbf{d}_2^*; \mathbf{r}_3, \mathbf{d}_3^*)$$
$$= \int d^3\mathbf{r}' G_{\alpha\beta}(\mathbf{r_p}, \mathbf{r}')\chi^{(3)}_{\gamma\delta\epsilon,\beta}G_{\gamma m}(\mathbf{r}', \mathbf{r_1})G_{\delta n}(\mathbf{r}', \mathbf{r_2})G_{\epsilon o}(\mathbf{r}', \mathbf{r_3})\mathbf{d}_{1m}^*\mathbf{d}_{2n}^*\mathbf{d}_{3o}^*. \tag{13}$$

We can further express the Green's function in its far-field form where $r \gg c/\omega$ and $r \gg r'$ as:

$$G_{\alpha,\beta}(\mathbf{r}, \mathbf{r_i}) = q_i^2 \frac{e^{iqr}}{r}g_{\alpha,\beta}(\frac{\mathbf{r}}{r}, \mathbf{r_i}), \tag{14}$$

where $g_{\alpha\beta}(\frac{\mathbf{r}}{r}, \mathbf{r}') = g_{\alpha\beta}(\mathbf{k}, \mathbf{r}')$ is a dimensionless scattering amplitude describing the conversion between the near-field point $r'$ and the plane wave propagating the in the direction $\mathbf{r}/r \parallel \mathbf{k}$. With such definition and by utilizing the reciprocity property:

$$G_{\alpha\beta}(\mathbf{r_1}, \mathbf{r_2}) = G_{\beta\alpha}(\mathbf{r_2}, \mathbf{r_1}), \tag{15}$$

we can rewrite Equation 13 as:

$$E_\alpha^{TSFG}(\mathbf{r}_p) = \frac{q_1^2 q_2^2 q_3^2 q_p^2}{r_1 r_2 r_3 r_p} e^{i(q_1 r_1 + q_2 r_2 + q_3 r_3 + q_p r_p)}$$
$$\int d^3 \mathbf{r}' g_{\alpha\beta}(\mathbf{k_p}, \mathbf{r}') \chi^{(3)}_{\gamma\delta\epsilon,\beta} g_{m\gamma}(\mathbf{k_1}, \mathbf{r}') g_{n\delta}(\mathbf{k_2}, \mathbf{r}') g_{o\epsilon}(\mathbf{k_3}, \mathbf{r}') d^*_{1m} d^*_{2n} d^*_{3o} \quad (16)$$

We now define the differential TSFG efficiency over the solid angle $d\Omega_p$ as:

$$d\Xi^{TSFG}(-\mathbf{k_p}, \mathbf{e_p^*}) = r_p^2 d\Omega_p \frac{\Phi_p(-\mathbf{k_p}, \mathbf{e_p^*})}{\Phi_1(-\mathbf{k_1}, \mathbf{e_1^*})\Phi_2(-\mathbf{k_2}, \mathbf{e_2^*})\Phi_3(-\mathbf{k_3}, \mathbf{e_3^*})}. \quad (17)$$

Here, $\mathbf{e_i} = \mathbf{d_i}/|\mathbf{d_i}|$ is the unit vector in the direction of the dipole source.

We now evaluate the differential TSFG in terms of the interacting fields $\mathbf{E}_i$, and with the help of Equation 12 and 16, we arrive at:

$$\frac{d\Xi^{TSFG}(-\mathbf{k_p}, \mathbf{e_p^*})}{d\Omega_p} = \left(\frac{2\pi}{c}\right)^2 q_p^4 \times$$
$$\left| \int d^3\mathbf{r}' e_{p\alpha} g_{\alpha\beta}(\mathbf{k_p}, \mathbf{r}') \chi^{(3)}_{\gamma\delta\epsilon,\beta} g_{m\gamma}(\mathbf{k_1}, \mathbf{r}') g_{n\delta}(\mathbf{k_2}, \mathbf{r}') g_{o\epsilon}(\mathbf{k_3}, \mathbf{r}') e^*_{1m} e^*_{2n} e^*_{3o} \right|^2. \quad (18)$$

## 1.3 Photon-Triplet Wavefunction

The complex wavefunction of a photon-triplet state generated via TOSPDC induced by the pump field $E_p$ has an amplitude given by:[S1]

$$\Psi(\mathbf{r_1}, \mathbf{r_2}, \mathbf{r_3}, \omega_1, \omega_2, \omega_3) = \int d^3\mathbf{r_0} G_{\sigma_1\alpha}(\mathbf{r_1}, \mathbf{r_0}) G_{\sigma_2\beta}(\mathbf{r_2}, \mathbf{r_0}) G_{\sigma_3\gamma}(\mathbf{r_3}, \mathbf{r_0}) \chi^{(3)}_{\alpha\beta\gamma\delta}(\mathbf{r_0}) E_{p\delta}(\mathbf{r_0}). \quad (19)$$

By similarly applying the far-field approximation to the Green's function and rearranging

terms according to the Green's reciprocity, Equation 19 may now take the form:

$$\Psi(\mathbf{r_1}, \mathbf{r_2}, \mathbf{r_3}, \omega_1, \omega_2, \omega_3) = \frac{q_1^2 q_2^2 q_3^2 q_p^2}{r_1 r_2 r_3 r_p} e^{i(q_1 r_1 + q_2 r_2 + q_3 r_3 + q_p r_p)} \times$$

$$\int d^3 r' g_{\sigma_1 m}(\mathbf{r'}, \mathbf{k_1}) g_{\sigma_2 n}(\mathbf{r'}, \mathbf{k_2}) g_{\sigma_3 o}(\mathbf{r'}, \mathbf{k_3}) \chi^{(3)}_{\gamma\delta\epsilon,\beta} g_{\alpha\beta}(\mathbf{k_p}, \mathbf{r'}) \mathbf{d}^*_{p\alpha} \quad (20)$$

Equation 20 resembles Equation 16 and comparing the two, we get another form of the general Lorentz reciprocity (Equation 8) derived earlier, as:

$$\Psi(\mathbf{r_1}, \mathbf{r_2}, \mathbf{r_3}, \omega_1, \omega_2, \omega_3) \mathbf{d}^*_{1m} \mathbf{d}^*_{2n} \mathbf{d}^*_{3o} = E^{TSFG}_\alpha(\mathbf{r}_p, \omega_p) \mathbf{d}^*_{p\alpha} \quad (21)$$

## 1.4 Photon Triplet Generation Rate

From the establishment of the third-order reciprocity relations, we now derive the photon triplet generation rate using Fermi's golden rule. The photon triplet generation rate is formally defined as the number of triplets per unit of its output energy $\hbar\omega_{i=1,2,3}$ over solid angle $d\Omega_{i=1,2,3}$ per unit time $t$:

$$\frac{1}{\hbar^3} \frac{dN_{\text{triplet}}}{dt d\omega_1 d\omega_2 d\omega_3 d\Omega_1 d\Omega_2 d\Omega_3} = \frac{W_{123}}{dQE_1 dQE_2 dQE_3}. \quad (22)$$

Here, $W_{123}$ is the transition rate calculated from the three-photon wavefunction (Equation 20). The denominator $dQE_{i=1,2,3}$ is the normalization over the quantum efficiency of an idealized photodetector modelled as a two-level system with dipole momenta matrix $\mathbf{d}_{i=1,2,3}$ and energies $\hbar\omega_{i=1,2,3}$. The uncalibrated transition rate $W_{123}$ and the quantum efficiency $dQE_{i=1,2,3}$ take the form:

$$W_{123} = \frac{2\pi}{\hbar} \delta(\hbar\omega_p - \hbar\omega_1 - \hbar\omega_2 - \hbar\omega_3) \left| \sum_{m,n,o} d^*_{1m} d^*_{2n} d^*_{3o} \Psi(\mathbf{r_1} m, \mathbf{r_2} n, \mathbf{r_3} o \leftarrow \mathbf{r_p}, \mathbf{e_p}) \right|^2, \quad (23)$$

$$\frac{dQE_i}{d\Omega_i} = \frac{4\pi^2 \omega_i \delta(\hbar\omega_p - \hbar\omega_i)}{c} \frac{|d_i|^2}{r^2 d\Omega_i} \quad i = 1, 2, 3. \quad (24)$$

We then substitute Equation 20 to Equation 23, rearrange the terms, and compare to Equation 18. This will cancel the integrals and we are left with constants, which upon simplification, leads to:

$$\frac{dN_{\text{triplet}}}{dt d\omega_1 d\omega_2 d\omega_3 d\Omega_1 d\Omega_2 d\Omega_3} = \frac{\hbar c}{2\pi} \frac{\lambda_p^4}{\lambda_1^3 \lambda_2^3 \lambda_3^3} \delta(\omega_p - \omega_1 - \omega_2 - \omega_3) \Phi_p \frac{d\Xi^{\text{TSFG}}}{d\Omega_p}. \qquad (25)$$

Equation 25 is the central result of the TOSPDC-TSFG correspondence valid for any arbitrary localized $\chi^{(3)}$ nonlinear system. We then integrate the equation over the output photon bandwidth and solid angles to express the terms in experimentally accessible variables. This leads to:

$$\frac{1}{\Phi_p} \frac{dN_{\text{triplet}}}{dt} = \frac{\hbar c}{2\pi} \frac{\lambda_p^4}{\lambda_1^3 \lambda_2^3 \lambda_3^3} \left[ \frac{4}{3} \left( \frac{2\pi c \Delta \lambda_i}{\lambda_i^2} \right)^2 \right] \left[ 4\pi^2 \Xi^{\text{TSFG}} \right], \qquad (26)$$

which is equivalent to Equation 3 shown in the article.

## 2 Finite element simulation details

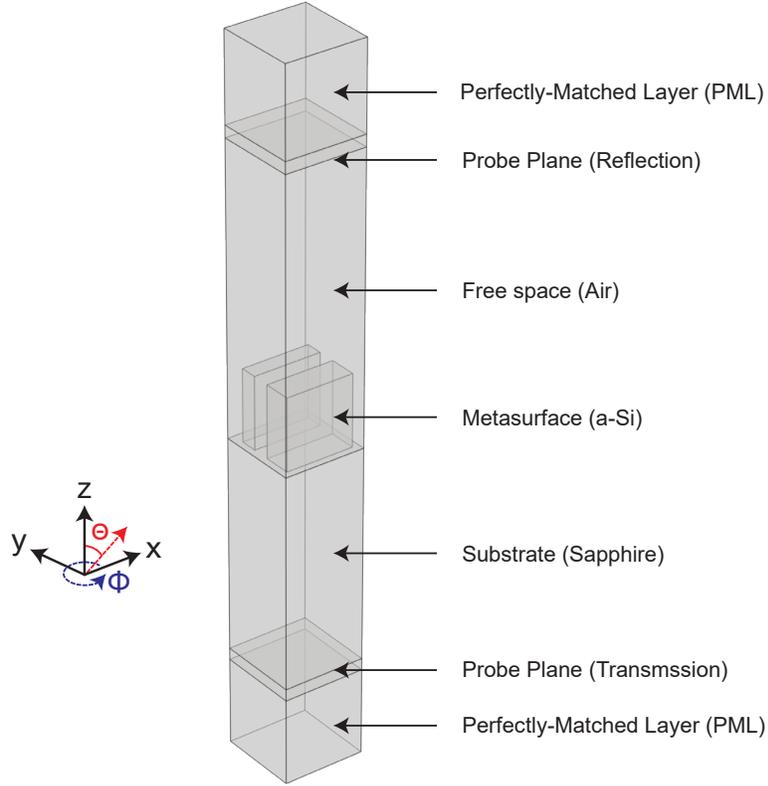

Figure S1: The simulation cell.

Full-wave frequency-domain electromagnetic simulations were performed via finite-element methods in COMSOL, which consists of a simulation cell shown in Figure S1, and is periodic along $x$ and $y$ directions.

In optimizing the metasurface design, we considered collinear degenerate three-wave SFG which is equivalent to third-harmonic generation (THG). Thus, a single pump is sufficient to generate the nonlinear field according to:[S3]

$$\mathbf{P}^{(3)}(3\omega) = \epsilon_0 \chi^{(3)}_{ijkl} \mathbf{E}_1^3(\omega). \qquad (27)$$

A plane wave with frequency $\omega$, wavevector $\mathbf{k} \parallel -\hat{\mathbf{z}}$, polarization $\mathbf{p} \parallel \mathbf{x}$, and magnitude $E_0$, i.e. a normally-incident TM plane wave, illuminates the sample. The scattered field are then probed along the planes above and below the samples, which constitute the reflected and transmitted linear response, respectively. The nonlinear response is then calculated according to the induced third-order polarizability (Equation 27).

Amorphous silicon (a-Si) was chosen as the nonlinear material for the metasurface, with sapphire ($\alpha$-Al$_2$O$_3$) as the substrate. The optical properties of silicon were obtained from ellipsometric measurements, with an additional emulated nonradiative loss factor $Q_{nr} = 150$ introduced as:

$$k_{\text{net}} = k_{\text{Si}}(\omega) + k', \quad k' = \frac{n(\omega)}{2Q_{nr}}. \tag{28}$$

This value of $Q_{nr}$ is adapted from previous experimental work on similar metasurface structure.[S4,S5] The optical properties of sapphire were taken from literature[S6] and were assumed to be lossless due to its extremely low extinction coefficient in the visible to near-infrared (NIR) range, which is the spectral region of interest.

The third-order susceptibility of silicon were taken to be $\chi^{(3)}_{xxxx} = \chi^{(3)}_{yyyy} = \chi^{(3)}_{zzzz} = 2.45 \times 10^{-19}$ m$^2$V$^{-2}$. For THG, only these three tensor components contribute to the nonlinear response.

Simulations were then performed at various bar widths $S_1$ and wavelengths $\lambda$. The transverse components of the scattered pump and nonlinear fields were calculated along the probe planes which correspond to the far-field response. The near-field enhancement was also calculated by extracting the maximum values of the pump and nonlinear fields within the metasurface volume.

For non-collinear, degenerate three-wave sum-frequency generation (TSFG), the model was extended to include three independent pump plane waves, each with a distinct wavevector $\mathbf{k_i}, (i = 1, 2, 3)$. This difference requires the full tensorial treatment for the calculation of

the nonlinear polarization according to:[S3]

$$P_i^{(3)}(\omega_p) = 6\epsilon_0 \sum_{j,k,l} \chi_{ijkl}^{(3)} E_j(\omega_1) E_k(\omega_2) E_l(\omega_3). \tag{29}$$

Explicitly, the Cartesian components of the third-order polarizability are listed in Table S1. Additionally, due to the isotropic crystallinity of amorphous silicon, the susceptibility tensor reduces to only 21 non-zero elements with only 3 independent components.[S3] They are:

$$yyzz = zzyy = zzxx = xxzz = xxyy = yyxx, \tag{30}$$

$$yzyz = zyzy = zxzx = xzxz = xyxy = yxyx, \tag{31}$$

$$yzzy = zyyz = zxxz = xzzx = xyyx = yxxy, \tag{32}$$

$$xxxx = yyyy = zzzz = xxyy + xyxy + xyyx. \tag{33}$$

This simplifies the calculation by only including the contribution of the components written in bold letters in Table S1.

By considering three-way symmetric divergence for the TOSPDC photons, the wavevectors of the TSFG waves are taken as:

$$\mathbf{k_1}(\theta, \phi_0) = k_{\lambda 1} \left[ \sin(\theta) \cos(\phi_0) \hat{\mathbf{x}} + \sin(\theta) \sin(\phi_0) \hat{\mathbf{y}} + \cos(\theta) \hat{\mathbf{z}} \right], \tag{34}$$

$$\mathbf{k_2}(\theta, \phi_0) = k_{\lambda 2} \left[ \sin(\theta) \cos(\phi_0 - 120°) \hat{\mathbf{x}} + \sin(\theta) \sin(\phi_0) \hat{\mathbf{y}} + \cos(\theta) \hat{\mathbf{z}} \right], \tag{35}$$

$$\mathbf{k_3}(\theta, \phi_0) = k_{\lambda 3} \left[ \sin(\theta) \cos(\phi_0 + 120°) \hat{\mathbf{x}} + \sin(\theta) \sin(\phi_0) \hat{\mathbf{y}} + \cos(\theta) \hat{\mathbf{z}} \right], \tag{36}$$

where $k_{\lambda_i} = \frac{2\pi n(\lambda_i)}{\lambda_i}$. The simulation was then swept over the polar angle $\theta = [0°, 90°]$ and azimuthal angles $\phi_0 = [0°, 30°]$, and the corresponding linear and nonlinear responses were calculated.

Table S1: Explicit expressions for the Cartesian components of the third-order polarizability.

| Index | X-component | Y-component | Z-component |
|---|---|---|---|
| 1 | **xxxxE1xE2xE3x** | yxxxE1xE2xE3x | zxxxE1xE2xE3x |
| 2 | xxxyE1xE2xE3y | **yxxyE1xE2xE3y** | zxxyE1xE2xE3y |
| 3 | xxxzE1xE2xE3z | yxxzE1xE2xE3z | **zxxzE1xE2xE3z** |
| 4 | xxyxE1xE2yE3x | **yxyxE1xE2yE3x** | zxyxE1xE2yE3x |
| 5 | **xxyyE1xE2yE3y** | yxyyE1xE2yE3y | zxyyE1xE2yE3y |
| 6 | xxyzE1xE2yE3z | yxyzE1xE2yE3z | zxyzE1xE2yE3z |
| 7 | xxzxE1xE2zE3x | yxzxE1xE2zE3x | **zxzxE1xE2zE3x** |
| 8 | xxzyE1xE2zE3y | yxzyE1xE2zE3y | zxzyE1xE2zE3y |
| 9 | **xxzzE1xE2zE3z** | yxzzE1xE2zE3z | zxzzE1xE2zE3z |
| 10 | xyxxE1yE2xE3x | **yyxxE1yE2xE3x** | zyxxE1yE2xE3x |
| 11 | **xyxyE1yE2xE3y** | yyxyE1yE2xE3y | zyxyE1yE2xE3y |
| 12 | xyxzE1yE2xE3z | yyxzE1yE2xE3z | zyxzE1yE2xE3z |
| 13 | **xyyxE1yE2yE3x** | yyyxE1yE2yE3x | zyyxE1yE2yE3x |
| 14 | xyyyE1yE2yE3y | **yyyyE1yE2yE3y** | zyyyE1yE2yE3y |
| 15 | xyyzE1yE2yE3z | yyyzE1yE2yE3z | **zyyzE1yE2yE3z** |
| 16 | xyzxE1yE2zE3x | yyzxE1yE2zE3x | zyzxE1yE2zE3x |
| 17 | xyzyE1yE2zE3y | yyzyE1yE2zE3y | **zyzyE1yE2zE3y** |
| 18 | xyzzE1yE2zE3z | **yyzzE1yE2zE3z** | zyzzE1yE2zE3z |
| 19 | xzxxE1zE2xE3x | yzxxE1zE2xE3x | **zzxxE1zE2xE3x** |
| 20 | xzxyE1zE2xE3y | yzxyE1zE2xE3y | zzxyE1zE2xE3y |
| 21 | **xzxzE1zE2xE3z** | yzxzE1zE2xE3z | zzxzE1zE2xE3z |
| 22 | xzyxE1zE2yE3x | yzyxE1zE2yE3x | zzyxE1zE2yE3x |
| 23 | xzyyE1zE2yE3y | yzyyE1zE2yE3y | **zzyyE1zE2yE3y** |
| 24 | xzyzE1zE2yE3z | **yzyzE1zE2yE3z** | zzyzE1zE2yE3z |
| 25 | **xzzxE1zE2zE3x** | yzzxE1zE2zE3x | zzzxE1zE2zE3x |
| 26 | xzzyE1zE2zE3y | **yzzyE1zE2zE3y** | zzzyE1zE2zE3y |
| 27 | xzzzE1zE2zE3z | yzzzE1zE2zE3z | **zzzzE1zE2zE3z** |

# 3 Linear and nonlinear properties of the metasurface

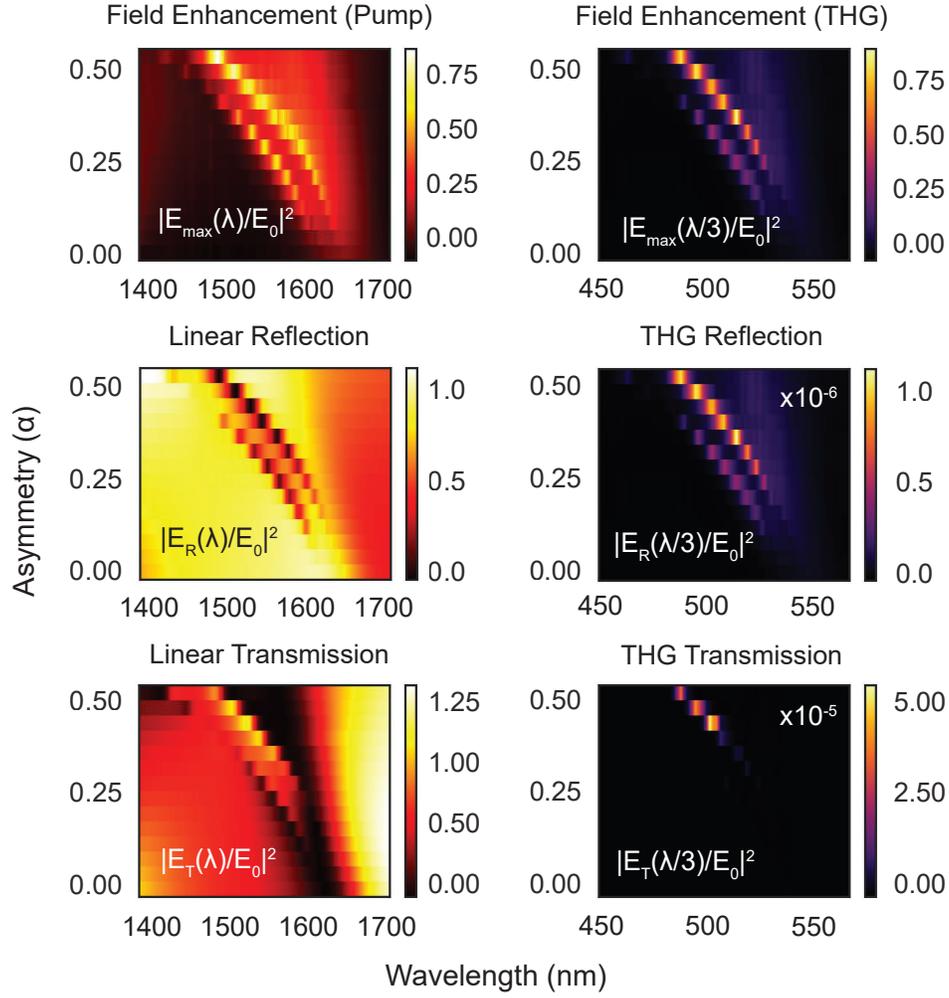

Figure S2: (Top panels) Field enhancement calculated as the maximum electric field within the meta-atom in the pump and TSFG (THG) regimes. (Middle and Bottom panels) Normalized reflection and transmission of the metasurface calculated as the average scattered electric fields along the top and bottom probe planes, respectively. Total radiated THG field at the bottom (forward) and top (backward) probe planes.

# 4 Critical Coupling Theoretical Fit

## 4.1 THG Enhancement

In Figure 2D of the main text, the THG enhancement was fitted with the equation below:

$$\frac{I(3\omega)}{I(\omega)^3} \propto Q_{nr}^3 \left[\frac{\alpha/\alpha_{cr}}{\alpha/\alpha_{cr} + 1}\right]^6, \tag{37}$$

where $I$ is the intensity of the field and $\alpha_{cr}$ is the critical asymmetry.[S7] The fitting was used to determine $\alpha_{cr}$ which resulted to 0.41.

## 4.2 Quality Factor

In Figure 2D, the theoretical fit to the quality factor is provided by:[S7]

$$Q_{tot}(\alpha) = \frac{Q_{nr}}{\alpha^2/\alpha_{cr}^2 + 1}, \tag{38}$$

where we used $\alpha_{cr} = 0.41$ from the previous fitting. This led to $Q(\alpha = 0) = Q_{nr} = 146$.

# 5 Non-collinear TSFG simulation results

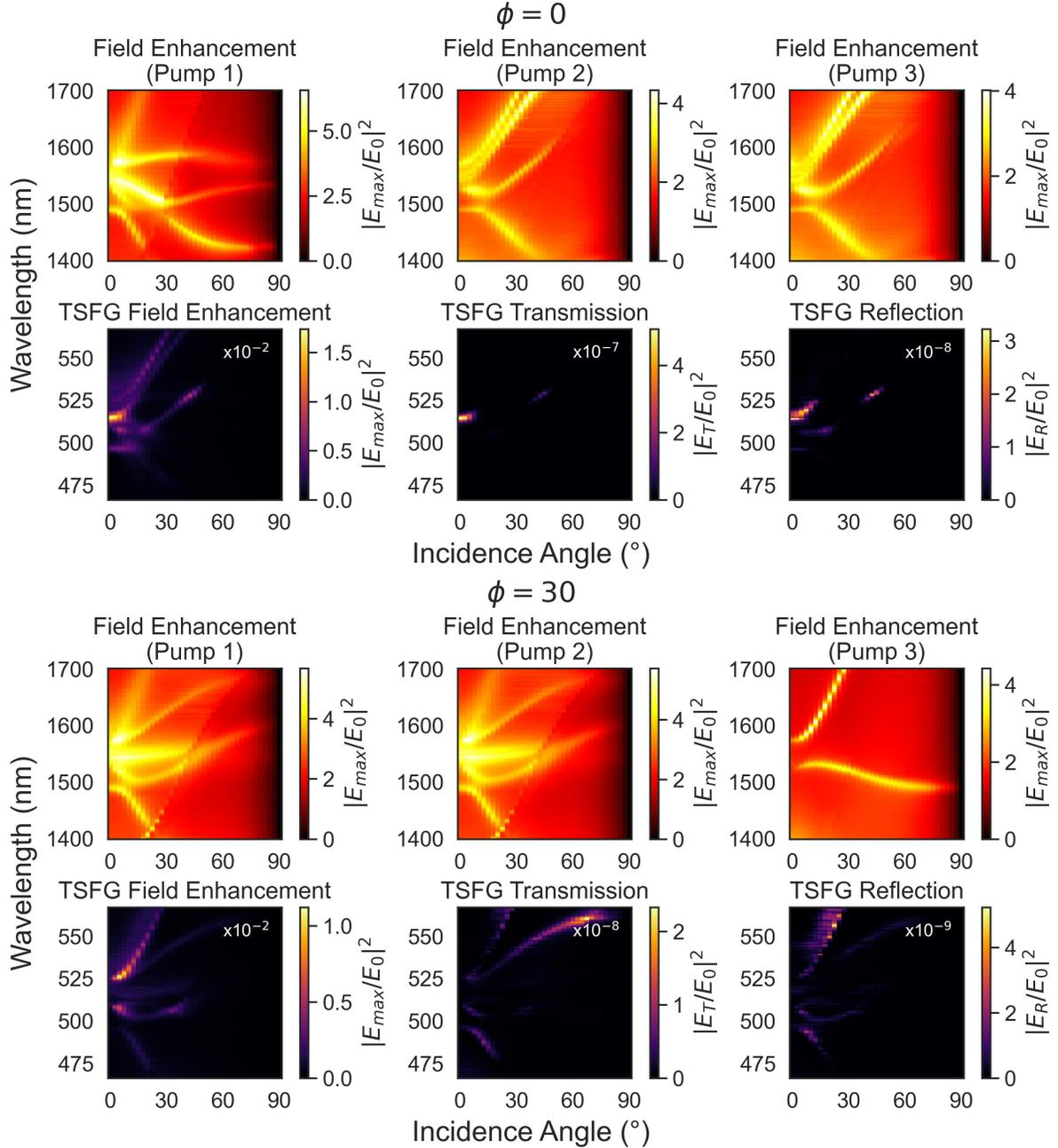

Figure S3: Results of the non-collinear TSFG simulation at $\phi = 0$ and $\phi = 30$. The upper panel shows the field-enhancement in the pump regime excited at three equidistant azimuth directions i.e. separated by 120°. The lower left panel shows the nonlinear TSFG field enhancement in the near-field. The lower center and lower right panels show the simulated transmitted and reflected TSFG response in the far-field, respectively.

# 6 TOSPDC Rate Calculation

To estimate the TOSPDC rate, we considered the pump flux power achievable from a focused beam using parameters of typical femtosecond lasers. The peak power of a focused laser pulse is calculated as:

$$P_{\text{peak}}(\text{W/m}^2) = \frac{P_{\text{ave}}(\text{W})}{\text{Repetition Rate (Hz)} \times \text{Pulse Width (s)} \times \text{Beam Area (m}^2)}. \tag{39}$$

Figure S4 shows the calculated peak power of a pulsed laser with an 80 MHz repetition rate and a 100 fs pulse width, focused by an objective to a spot size of 1–3 µm in diameter.

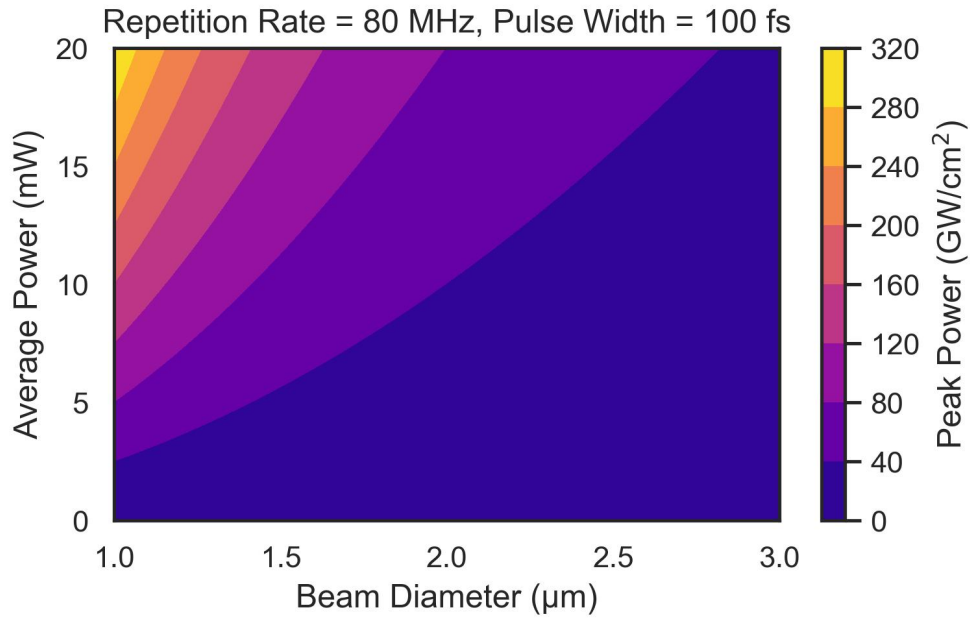

Figure S4: Calculated peak power of a focused beam from typical femtosecond laser specifications.

In the text, we used $\Phi_p = 40$ GWcm$^{-2}$ for the pump power which is achievable in a pulsed laser system described above with average power $P_{\text{ave}} = 10$ mW focused to a spot size with 2 um diameter.



\end{document}